\documentclass[useAMS,usenatbib]{mn2e}
\usepackage{color}
\usepackage{graphicx}
\usepackage{url}

\title[GWs from Eccentric Neutron Star Binaries]{Stochastic Background of Gravitational Waves Generated by Eccentric Neutron Star Binaries}
\author[Edgard F. D. Evangelista and Jos\'{e} C. N. de Araujo]{Edgard F. D. Evangelista$^{1}$\thanks{E-mails: edgard.evangelista@inpe.br; jcarlos.dearaujo@inpe.br} and  Jos\'{e} C. N. de Araujo$^{1}$\footnotemark[1]\\
$^{1}$Instituto Nacional de Pesquisas Espaciais, Divis\~{a}o de Astrof\'{i}sica, Av. dos Astronautas 1758, S J dos Campos, 12227-010 SP, Brazil}
\begin{document}

\date{}

\pagerange{\pageref{firstpage}--\pageref{lastpage}} \pubyear{2015}

\maketitle

\label{firstpage}

\begin{abstract}
Binary systems emit gravitational waves in a well-known pattern; for binaries in circular orbits, the emitted radiation has a frequency that is twice the orbital frequency. Systems in eccentric orbits, however, emit gravitational radiation in the higher harmonics too. In this paper, we are concerned with the stochastic background of gravitational waves generated by double neutron star systems of cosmological origin in eccentric orbits. We consider in particular the long-lived systems, that is, those binaries for which the time to coalescence is longer than the Hubble time ($\sim 10$Gyr). Thus, we consider double neutron stars with orbital frequencies ranging from $10^{-8}$ to $2\times 10^{-6}$Hz. Although in the literature some papers consider the spectra generated by eccentric binaries, there is still space for alternative approaches for the calculation of the backgrounds. In this paper, we use a method that consists in summing the spectra that would be generated by each harmonic separately in order to obtain the total background. This method allows us to clearly obtain the influence of each harmonic on the spectra. In addition, we consider different distribution functions for the eccentricities in order to investigate their effects on the background of gravitational waves generated. At last, we briefly discuss the detectability of this background by space-based gravitational wave antennas and pulsar timing arrays.
\end{abstract}


\section{Introduction}
\label{sec1}
It is well known that binary systems emit gravitational waves (GWs) and, for circular orbits, the frequency of the emitted radiation is twice the orbital frequency. On the other hand, for eccentric orbits, the pattern of emission has a more complex form: the emitted signals are also formed by the higher harmonics of the orbital frequency. In fact, the higher the orbital eccentricity is, the greater the contribution of the higher harmonics to the emitted radiation \citep[see e.g.,][]{peters,thorne}.

Since many binaries can have non-negligible orbital eccentricities \citep{kowalska2}, it is interesting to investigate the behaviour of the spectra when eccentricity is taken into account. Particularly, this paper is  addressed to the stochastic background of GWs generated by long-lived double neutron star (DNS) systems of cosmological origin in eccentric orbits. We consider that a system is long-lived if its lifetime is longer than the  Hubble time, namely $\sim 10$ Gyr \citep{kowalska2}. Quantitatively, this assumption implies that the binaries have orbital frequencies in the range $10^{-8}-2\times 10^{-6}$Hz, as explained in Section~\ref{sec5}. We should bear in mind that in this case we can neglect the time evolution of the orbital parameters of the systems.

It is worth noting that the inclusion of eccentricity in the calculation of stochastic backgrounds generated by compact binaries is not new. In fact, one can find in the literature interesting papers on the stochastic backgrounds generated by eccentric binaries, such as \citet{ignatiev}, \citet{kowalska2} and \citet{enoki}. In \citet{ignatiev}, the authors study, for example, the GW background produced by coalescing neutron stars in the Galaxy in eccentric orbits; in \citet{kowalska2} the authors study the backgrounds generated by Population III binaries, giving special attention to the spectra generated by long-lived and short-lived binaries; in \citet{enoki}, the authors studied the background generated by supermassive black hole binaries.

In general, the procedure to obtain the backgrounds generated by eccentric binaries includes, as a key component, the spectra of energy emitted by the source, which is written as $dE/d\nu$, where $E$ is the energy and $\nu$ is, in our case, the orbital frequency. In the literature, the authors usually handle such a quantity following the steps: first, they sum the spectra $dE/d\nu_{n}$ corresponding to each harmonic, where $n$ is the number of the harmonic, in order to obtain the total spectrum $dE/d\nu$; subsequently, $dE/d\nu$ is included in the formula that determines the stochastic background. Here we use a different method: we calculate the backgrounds that would be generated by each harmonic of the emitted frequency and then we sum all these signals in order to obtain the total background. As we will see, this method allows us to clearly analyse the influence of each harmonic on the resulting backgrounds.

In addition, we consider in our calculations different distribution functions for the orbital eccentricity $\varepsilon$, and study the influence of such distributions on the resulting backgrounds. Particularly, we considered the case in which the distribution function of the eccentricities has the form of a Dirac delta function, that is, all the systems have the same eccentricity. We consider scenarios in which all systems have eccentricities of 0.25, 0.5, 0.75 and 0.9. Moreover, we consider the case of circular orbits (null eccentricity) and also another distribution that covers a wide range of eccentricities, namely, a uniform distribution ranging from $\varepsilon=0$ to $\varepsilon=0.9$.

The paper is organized as follows: in Section~\ref{sec2}, we derive the fundamental equations used to calculate the stochastic background of GWs generated by DNS systems of cosmological origin in eccentric orbits; in Section~\ref{sec3}, we calculate the spectrum of energy for eccentric systems; in Section \ref{sec4}, we present and discuss the distribution functions of the eccentricities $f(\varepsilon)$ adopted; in Section~\ref{sec5}, we discuss the maximum and minimum frequencies of the spectra and the population characteristics of the systems, such as the star formation rate density $\dot{\rho_{\ast}}(z)$ and the mass fraction $\lambda_{\rmn{dns}}$ of stars that is converted in DNSs; we present the results in Section~\ref{sec6} and the conclusions and perspectives in Section~\ref{sec7}.

\section{Fundamental Equation}
\label{sec2}

For the calculation of the spectra generated by each harmonic, we follow the method described in \citet{regimbau2} and \citet{regimbau1}, who calculated the backgrounds in terms of the energy density parameter $\Omega_{\rmn{gw}}$, namely

\begin{equation}
 \Omega_{\rmn{gw}}(\nu_{0})=\frac{1}{\rho_{\rmn{c}}c^{3}}\nu_{0}F(\nu_{0}) ,
\label{eq1}
\end{equation}
where $\rho_{\rmn{c}}=3H_{0}^{2}/8\pi G$ is the critical density of the universe, $H_{0}=70\rmn{km}\hspace{3pt}\rmn{s}^{-1}\rmn{Mpc}^{-1}$ is the Hubble parameter, $\nu_{0}$ is the observed frequency and $F(\nu_{0})$ is the spectral density of the flux, which is given by \citep{ferrari1,ferrari2}

\begin{equation}
\label{eq2}
 F(\nu_{0})=\int f(\nu_{0},z)\frac{dR}{dz}dz ,
\end{equation}
where $z$ is the redshift and $f(\nu_{0},z)$ is the energy flux per unit frequency generated by a unique source, which has the form \citep{regimbau2}

\begin{equation}
\label{eq3}
 f(\nu_{0},z)=\frac{1}{4\pi r(z)^{2}}\left.\frac{dE}{d\nu}\right|_{\nu=\nu_{0}(1+z)}.
\end{equation}

In equation~(\ref{eq3}), $r(z)$ is the comoving distance, $\nu=\nu_{0}(1+z)$ is the emitted frequency and $dE/d\nu$ is the spectrum of energy emitted by a binary system, calculated in the source frame. In addition, in equation~(\ref{eq2}), the event rate $dR/dz$ is calculated by

\begin{equation}
 \label{eq4}
 \frac{dR}{dz}=\lambda_{\mbox{\scriptsize{dns}}}\frac{\dot{\rho_{\ast}}(z)}{1+z}
\frac{dV}{dz} ,
\end{equation}
where $\dot{\rho_{\ast}}(z)$ is the star formation rate density (SFRD), $\lambda_{\rmn{dns}}$ is the mass fraction of stars that is converted into DNSs, and $dV/dz$ is given by

\begin{equation}
\label{eq5}
\frac{dV}{dz}=4\pi\left(\frac{c}{H_{0}}\right)\frac{r(z)^{2}}{\sqrt{\Omega_{\rmn{M}}(1+z)^{3}+\Omega_{\rmn{k}}(1+z)^{2}+\Omega_{\Lambda}}}.
\end{equation}
We use equation~(\ref{eq5}) in a scenario in which the density parameters obey the relations $\Omega_{\rmn{M}}=\Omega_{\rmn{DM}}+\Omega_{\rmn{B}}$ and $\Omega_{\rmn{k}}+\Omega_{\rmn{M}}+\Omega_{\Lambda}=1$, with the subindices M, DM, B, k and $\Lambda$ corresponding to matter, dark matter, baryonic matter, curvature and cosmological constant, respectively. Here, we consider a cosmological scenario in which $\Omega_{\rmn{M}}=0.3$, $\Omega_{\rmn{B}}=0.04$, $\Omega_{\rmn{k}}=0$ and $\Omega_{\Lambda}=0.7$.

In addition, the eccentricity is introduced in the calculation of the background by means of a distribution function $f(\varepsilon)$ in equation~(\ref{eq1}). So, with the above considerations and performing all the substitutions, equation~(\ref{eq1}) assumes the following form
\begin{eqnarray}\nonumber
 \label{eq6}
\lefteqn{\Omega_{\mbox{\scriptsize{gw}}}(\nu_{0},\varepsilon)=\frac{8\pi G}{3H_{0}^{3}c^{2}}\lambda_{\mbox{\scriptsize{dns}}}\nu_{0}}\\
&&\times\int^{z_{f}(\nu_{0})}_{z_{i}(\nu_{0})}\frac{\dot{\rho_{\ast}}(z)}{(1+z)E(z)}\left.\frac{dE}{d\nu}\right|_{\nu=\nu_{0}(1+z)}f(\varepsilon)dz,
\end{eqnarray}
where the meaning of the limits of integration, $z_{f}(\nu_{0})$ and $z_{i}(\nu_{o})$, will
be considered in Section \ref{sec5}.

\section[]{Energy Emitted by Eccentric Binaries}
\label{sec3}

In the previous section, we noticed that the spectrum of energy $dE/d\nu$ is a key ingredient for the calculation of the backgrounds, since it provides information concerning the patterns of emission of the systems as a function of the orbital frequency. We can now obtain an expression for the energy spectrum for eccentric systems starting from

\begin{equation}
 \label{eq7}
 \frac{dE}{d\nu_{n}}=\frac{dE}{dt}\frac{dt}{d\Omega}\frac{d\Omega}{d\nu_{n}} ,
\end{equation}
where we included the index $n$ in the notation, since equation~(\ref{eq7}) refers to the spectrum generated by each harmonic. In addition, $\Omega$ is the angular orbital frequency and $dt/d\Omega$ is given by

\begin{equation}
\label{eq8}
 \frac{d\Omega}{dt}=\frac{d\Omega}{da}\frac{da}{dt} ,
\end{equation}
in which '$a$' represents the orbital semimajor axis. In our case, $dE/dt$ and $da/dt$ are given by \citep{peters}

\begin{equation}
\label{eq9}
 \frac{dE}{dt}=\frac{32}{5}\frac{G^{4}}{c^{5}}\frac{m_{1}^{2}m_{2}^{2}(m_{1}+m_{2})}{a^{5}}g(n,\varepsilon);\\
\end{equation}
and
\begin{equation}
 \frac{da}{dt}=-\frac{64}{5}\frac{G^{3}m_{1}m_{2}(m_{1}+m_{2})}{c^{5}a^{3}}F(\varepsilon),
\end{equation}
where $m_{1}$ and $m_{2}$ are the masses of the components of the system. In addition, the functions $g(n,\varepsilon)$ and $F(\varepsilon)$ have the form
\begin{eqnarray}\nonumber
\label{exc1}
\lefteqn{g(n,\varepsilon)=\frac{n^4}{32}\left\{\left[J_{n-2}(n\varepsilon)-2\varepsilon J_{n-1}(n\varepsilon)+\frac{2}{n}J_{n}(n\varepsilon)+\right.\right.} + \\
&&\left.+2\varepsilon J_{n+1}(n\varepsilon)-J_{n+2}(n\varepsilon)\right]^2+(1-\varepsilon^2)\left[J_{n-2}(n\varepsilon)\right.+\nonumber + \\
&&\left.\left.-2\varepsilon J_{n}(n\varepsilon)+J_{n+2}(n\varepsilon)\right]^2+\frac{4}{3n^2}\left[J_{n}(n\varepsilon)\right]^2\right\},
\end{eqnarray}
where $J_{n}(n\varepsilon)$ are the Bessel functions of order $n$, and
\begin{equation}
\label{exc2}
F(\varepsilon)=\frac{1+73\varepsilon^{2}/24+37\varepsilon^{4}/96}{(1-\varepsilon^{2})^{7/2}}.
\end{equation}

Moreover, equations~(\ref{exc1})~and~(\ref{exc2}) are related to each other by
\begin{equation}
F(\varepsilon)=\sum_{n=2}^{\infty} g(n,\varepsilon),
\end{equation}
and $d\Omega/da$ is obtained from Kepler's third law.

Now, recall that systems in eccentric orbits emit radiation at harmonics of the orbital frequency, which is related to the angular orbital frequency by $\Omega=2\pi\nu_{\mbox{\scriptsize{orb}}}$. Thus, we can write the emitted frequency $\nu_{n}$ of the harmonic $n$ as a function of $n$ and $\Omega$, namely

\begin{equation}
 \label{eq10}
 \nu_{n}=\frac{n\Omega}{2\pi}\hspace{5pt} \mbox{for}\hspace{5pt} n\geq 2,
\end{equation}
which enable us to calculate $d\Omega/d\nu_{n}$ in equation~(\ref{eq7}).

Now, performing all the necessary substitutions, equation~(\ref{eq7}) is transformed into

\begin{equation}
\label{eq11}
\frac{dE}{d\nu_{n}}=\frac{1}{3}\left(\frac{2\pi G}{n}\right)^{2/3}\frac{m_{1}m_{2}}{(m_{1}+m_{2})^{1/3}}\nu_{n}^{-1/3}\frac{g(n,\varepsilon)}{F(\varepsilon)}.
\end{equation}

Note that for $n=2$, $F(\varepsilon)=1$ and $g(n,\varepsilon)=\delta_{n2}$, equation~(\ref{eq11}) comes down to the expression valid for circular orbits \citep[see e.g.,][]{ferrari3}.

In addition, $g(n,\varepsilon)/F(\varepsilon)$ has the following meaning: it provides the fraction of energy emitted at the harmonic $n$ by a system with orbital eccentricity $\varepsilon$.

\section[]{The Distribution Functions of the Eccentricities}
\label{sec4}

We are interested in studying the role of the distribution function of the orbital eccentricities on the stochastic background of GWs generated by DNS systems of cosmological origin. We consider distributions that, although could be considered unrealistic, can provide an insight into the influence of these functions on the spectra.

First, we consider a scenario where all the DNSs have the same orbital eccentricity, such that

\begin{equation}
\label{deltaEccentr}
f(\varepsilon)=\delta(\varepsilon-\varepsilon^{\prime}),
\end{equation}
where $\delta(\varepsilon-\varepsilon^{\prime})$ is a Dirac delta function. Particularly, we consider the cases for $\varepsilon^{\prime}=0$ (a population of circular DNSs), 0.25, 0.5, 0.75 and 0.9.

We also consider a distribution that covers a wide range of values of $\varepsilon$. For this purpose, we use a uniform and normalized function that is given by

\begin{equation}
f(\varepsilon)=\left\{\begin{array}{rc}
&(\varepsilon_{\mbox{\scriptsize{max}}}-\varepsilon_{\mbox{\scriptsize{min}}})^{-1}\hspace{7pt}\mbox{for} \hspace{7pt}\varepsilon_{\mbox{\scriptsize{min}}}\leq \varepsilon\leq \varepsilon_{\mbox{\scriptsize{max}}}\\
\vspace{2pt}\\
&0 \hspace{29pt}\mbox{otherwise},
\end{array}\right.
\label{eq16}
\end{equation}
where $\varepsilon_{\mbox{\scriptsize{min}}}=0$ and $\varepsilon_{\mbox{\scriptsize{max}}}=0.9$ are the minimum and the maximum eccentricities, respectively. This distribution provides an interesting case to be compared with the distribution functions given in terms of the Dirac delta functions.

\section[]{The Stochastic Background}
\label{sec5}

Before proceeding, we pay attention to the choice of the minimum and maximum values of the orbital frequencies used in the calculations of the backgrounds. First, as already mentioned, we consider in this paper long-lived DNSs, which lead us to the determination of a maximum value for the initial frequency. We mean by a 'long-lived system', a binary whose lifetime is at least of the order of the Hubble time, that is, $\sim 10$Gyr.

In addition, recall that the greater the orbital eccentricity is, the shorter is the binary system lifetime. Therefore, we choose the maximum frequency such that the most eccentric systems considered in this paper have lifetimes $T_{\mbox{\scriptsize{life}}}$ equal to the Hubble time. The expression for $T_{\mbox{\scriptsize{life}}}$, considering $\varepsilon$ close to $1$, is given by \citep{peters2}

\begin{equation}
\label{lifetime}
T_{\mbox{\scriptsize{life}}}\approx\frac{768}{425}\frac{a_{0}^{4}}{4\beta}(1-\varepsilon_{0}^{2})^{7/2},
\end{equation}
where $\beta=64G^{3}m_{1}m_{2}(m_{1}+m_{2})/5c^{5}$, $a_{0}$ is the initial orbital semimajor axis and $\varepsilon_{0}$ is the initial eccentricity. Considering, for example, $T_{\mbox{\scriptsize{life}}}\sim 10$Gyr, $m_{1}=m_{2}=1.4\mbox{M}_{\odot}$ and $\varepsilon_{0}=0.9$ in equation~(\ref{lifetime}) and using Kepler's third law, we obtain a frequency of $\approx 2\times 10^{-6}$Hz.

On the other hand, according to \citet{rosado}, there is a minimum frequency one must consider in the study of GWs generated by binary systems. He argues that the emitted radiation with frequencies lower than a given value should not be considered, since other mechanisms that take away energy from the systems are more effective (see \citealt{rosado} for a detailed discussion.)

Still following the discussion of \citet{rosado}, we should bear in mind two facts: it is difficult to define exactly the values of the minimum frequencies; on the other hand, such values are not important in practice, since the choice of a particular frequency will not have noticeable effects on the results. Considering these arguments, we adopted the rather arbitrary value of $10^{-8}$Hz in our calculations.

To proceed, the next step is to determine the SFRD $\dot{\rho_{\ast}}(z)$ and the value of $\lambda_{\mbox{\scriptsize{dns}}}$.

Concerning the SFRD, many different proposals are considered in the literature, although they do not differ from each other very significantly. Here we adopt, as a fiducial one, that given by \citet{sfrd}, namely

\begin{equation}
\label{eq12}
\dot{\rho}_{\ast}(z)=\rho_{m}\frac{\beta e^{\alpha(z-z_{m})}}{\beta-\alpha+\alpha e^{\beta(z-z_{m})}} ,
\end{equation}
where $\alpha=3/5$, $\beta=14/15$, $z_{m}=5.4$ and $\rho_{m}=0.15\mbox{M}_{\odot}\mbox{yr}^{-1}\mbox{Mpc}^{-3}$ fixing the normalization. It is worth mentioning that the authors considered a $\Lambda$CDM cosmology, where the density parameters have the values given in Section~\ref{sec2}. Moreover, they considered the Hubble parameter as $H_{0}=100\hspace{2pt}h\hspace{2pt}\mbox{km}\hspace{2pt}\mbox{s}^{-1}\hspace{1pt}\mbox{Mpc}^{-3}$ with $h=0.7$.

The mass fraction $\lambda_{\mbox{\scriptsize{dns}}}$ \cite[see e.g.,][ for a detailed discussion]{regimbau3} is given by

\begin{equation}\label{}
    \lambda_{\mbox{\scriptsize{dns}}}=\beta_{\mbox{\scriptsize{ns}}}f_{p}\Phi_{\mbox{\scriptsize{ns}}},
\end{equation}
where $\beta_{\mbox{\scriptsize{ns}}}$ is the fraction of binary systems that survive to the second supernova event, $f_{p}$ provides the fraction of massive binaries (that is, binary systems where both components could generate a supernova event) formed inside the whole population of stars, and $\Phi_{\mbox{\scriptsize{ns}}}$ is the mass fraction of neutron star progenitors. In our case, we assume that $\Phi_{\mbox{\scriptsize{ns}}}$ is calculated by
\begin{equation}\label{phi}
    \Phi_{\mbox{\scriptsize{ns}}}=\int^{25}_{8}\phi(m)dm ,
\end{equation}
where $\phi(m)$ is the Salpeter mass function \citep{salpeter}, with $\phi(m)=Am^{-(1+x)}$, $A=0.17$ and $x=1.35$. Moreover, note that we are considering that the progenitors of the neutron stars have masses in the range $8-25\mbox{M}_{\odot}$.

According to \citet{regimbau3}, the parameters $\beta_{\mbox{\scriptsize{ns}}}$ and $f_{p}$ depend on different assumptions. First, in the evolutionary scenario of massive binaries the authors considered that none of the stars underwent recycle accretion. On the other hand, the velocity distribution of the natal kick affects the parameter $\beta_{\mbox{\scriptsize{ns}}}$, since an imparted kick could disrupt binaries or, less likely, prevent the disruption of the system. Particularly, the authors adopted a one-dimensional velocity dispersion of $\approx 80\mbox{km s}^{-1}$.

Still according to \citet{regimbau3}, the estimates of $f_{p}$ depend on $\beta_{\mbox{\scriptsize{ns}}}$ itself and on the ratio between single and double NS systems in the Galaxy. As a result, there are uncertainties in the values of these parameters. \citet{regimbau3} then estimate the relative uncertainties in $f_{p}$ and $\beta_{\mbox{\scriptsize{ns}}}$. Numerically, they obtained $\sigma_{f_{p}}/f_{p}\approx 0.5$ and $\sigma_{\beta_{\mbox{\scriptsize{ns}}}}/\beta_{\mbox{\scriptsize{ns}}}\approx 0.75$, leading to a relative uncertain of $\sigma_{\lambda_{\mbox{\scriptsize{dns}}}}/\lambda_{\mbox{\scriptsize{dns}}}\approx 0.9$ in $\lambda_{\mbox{\scriptsize{dns}}}$. The authors stated that these uncertainties are solely formal, resulting from the simulations they performed, which depend on the evolutionary scenario for the progenitors.

Concerning the range of masses of the neutron star progenitors, recall that the minimum and maximum values are subject of discussions. However, the limits we adopt in equation~(\ref{phi}) are usual (see, e.g., \citealt{ferrari1, ferrari2}). We refer the reader, for example, to \citet{heger}, \citet{smartt} and \citet{carroll}, who discuss this subject in detail.

Numerically, we have $\beta_{\mbox{\scriptsize{ns}}}=0.024$, $f_{p}=0.136$, $\Phi_{\mbox{\scriptsize{ns}}}=5.97\times 10^{-3}\mbox{M}_{\odot}^{-1}$ and, consequently, $\lambda_{\mbox{\scriptsize{dns}}}=1.95\times 10^{-5}\mbox{M}_{\odot}^{-1}$.

For the limits of integration $z_{f}(\nu_{0})$ and $z_{i}(\nu_{0})$ in equation~(\ref{eq6}), we follow  \cite{regimbau2} and \cite{regimbau1}, namely

\begin{equation}
\label{eq13}
z_{f}(\nu_{0})=\left\{\begin{array}{rc}
z_{\mbox{\scriptsize{max}}} \hspace{5pt}\mbox{if} \hspace{10pt}\nu_{0}<\frac{\nu_{\mbox{\scriptsize{max}}}}{1+z_{\mbox{\scriptsize{max}}}} \\
\vspace{2pt}\\
\frac{\nu_{\mbox{\scriptsize{max}}}}{\nu_{0}}-1 \hspace{10pt}\mbox{otherwise} ;
\end{array}\right.
\end{equation}

and

\begin{equation}
\label{eq14}
z_{i}(\nu_{0})=\left\{\begin{array}{rc}
z_{\mbox{\scriptsize{min}}} \hspace{5pt}\mbox{if} \hspace{10pt}\nu_{0}>\frac{\nu_{\mbox{\scriptsize{min}}}}{1+z_{\mbox{\scriptsize{min}}}} \\
\vspace{2pt}\\
\frac{\nu_{\mbox{\scriptsize{min}}}}{\nu_{0}}-1 \hspace{10pt}\mbox{otherwise} ,
\end{array}\right.
\end{equation}
where $\nu_{\mbox{\scriptsize{min}}}=10^{-8}$Hz, $\nu_{\mbox{\scriptsize{max}}}=2\times 10^{-6}$Hz, $z_{\mbox{\scriptsize{min}}}=0$ and $z_{\mbox{\scriptsize{max}}}=20$ are the minimum emitted frequency, the maximum emitted frequency, the minimum redshift and the maximum redshift, respectively.

Using the ingredients shown above, integrating over the eccentricity $\varepsilon$ and substituting the numerical values, equation~(\ref{eq6}) takes the form

\begin{eqnarray}\nonumber
\label{eq17}
\lefteqn{\Omega_{\rmn{gw}}(n,\nu_{0})=1.28\times 10^{-10}\left(\frac{\nu_{0}}{n}\right)^{2/3}} \\
&&\times\int^{z_{f}(\nu_{0})}_{z_{i}(\nu_{0})}\frac{\mbox{e}^{0.6(z-5.4)}}{0.333+0.6\mbox{e}^{0.933(z-5.4)}}\frac{dz}{(1+z)^{4/3}E(z)}\nonumber\\
&&\times\int^{\varepsilon_{\mbox{\scriptsize{max}}}}_{\varepsilon_{\mbox{\scriptsize{min}}}}\frac{g(n,\varepsilon)f(\varepsilon)}{F(\varepsilon)}d\varepsilon,
\end{eqnarray}
where $E(z)=\sqrt{0.3(1+z)^{3}+0.7}$ and $f(\varepsilon)$ is given by either equation~(\ref{deltaEccentr}) or equation~(\ref{eq16}).

\section[]{Results}
\label{sec6}

\begin{figure}
\centering
\rotatebox{-90}{\includegraphics[width=60mm]{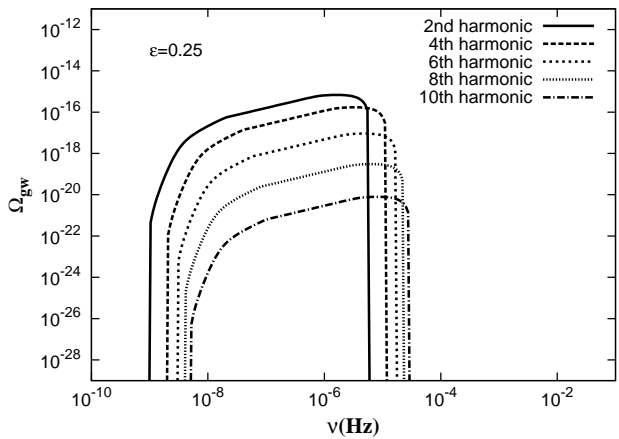}}
\caption{Backgrounds generated by some harmonics considering that all systems have eccentricity $0.25$.}
\label{fig1}
\end{figure}

\begin{figure}
\centering
\rotatebox{-90}{\includegraphics[width=60mm]{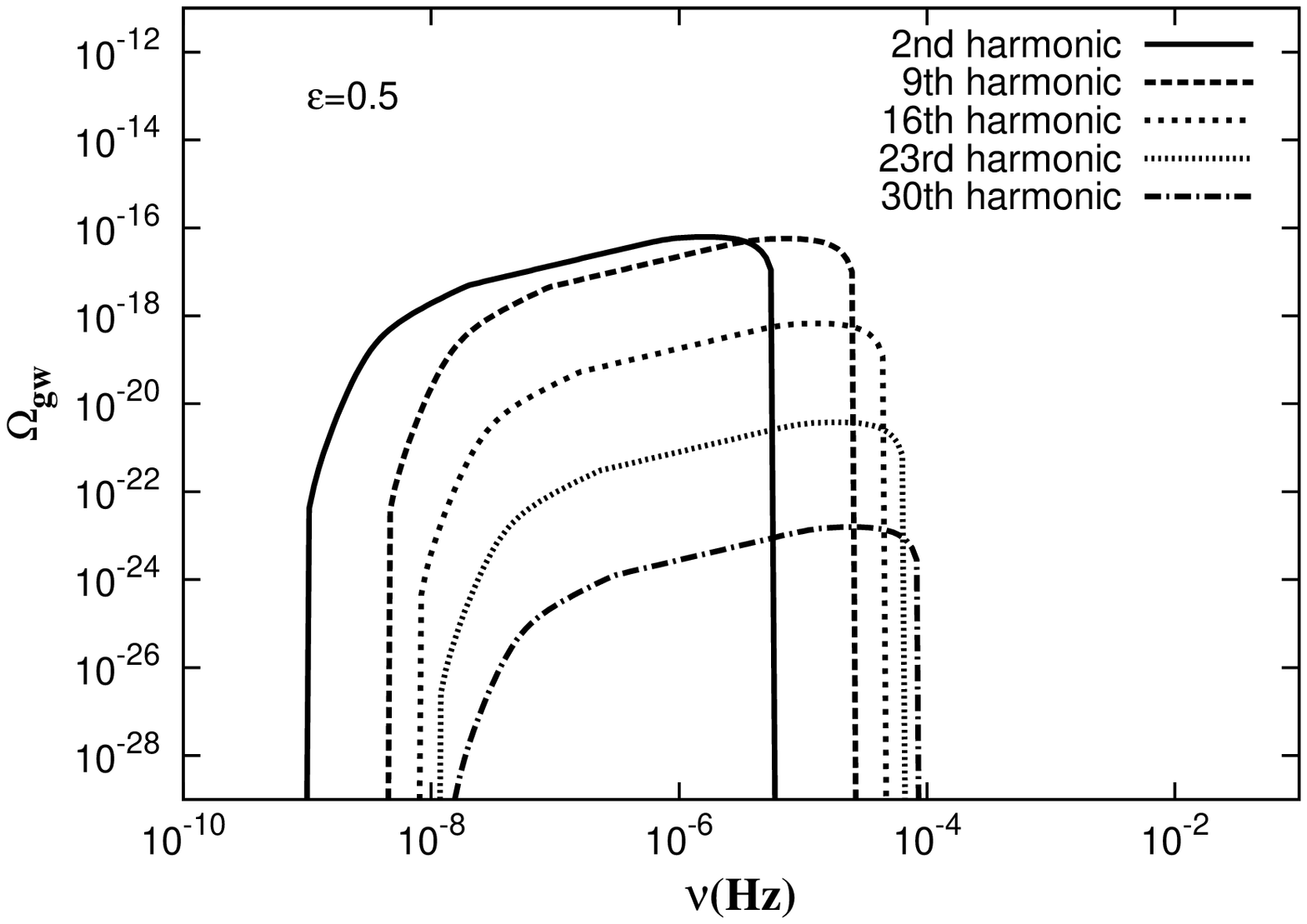}}
\caption{Same as in Fig. 1 for eccentricity $0.5$.}
\label{fig2}
\end{figure}

\begin{figure}
\centering
\rotatebox{-90}{\includegraphics[width=60mm]{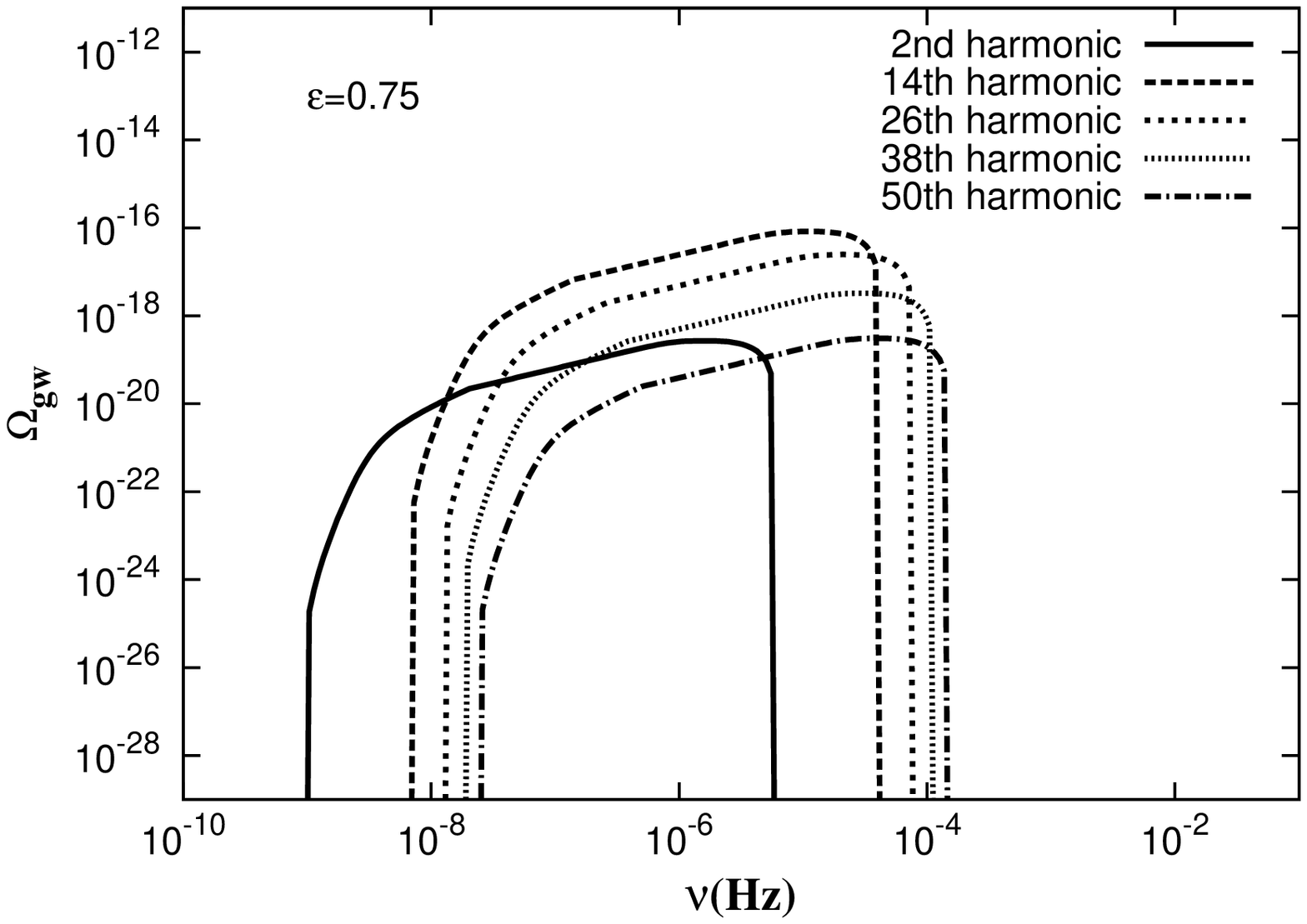}}
\caption{Same as in Fig. 1 for eccentricity $0.75$.}
\label{fig3}
\end{figure}

\begin{figure}
\centering
\rotatebox{-90}{\includegraphics[width=60mm]{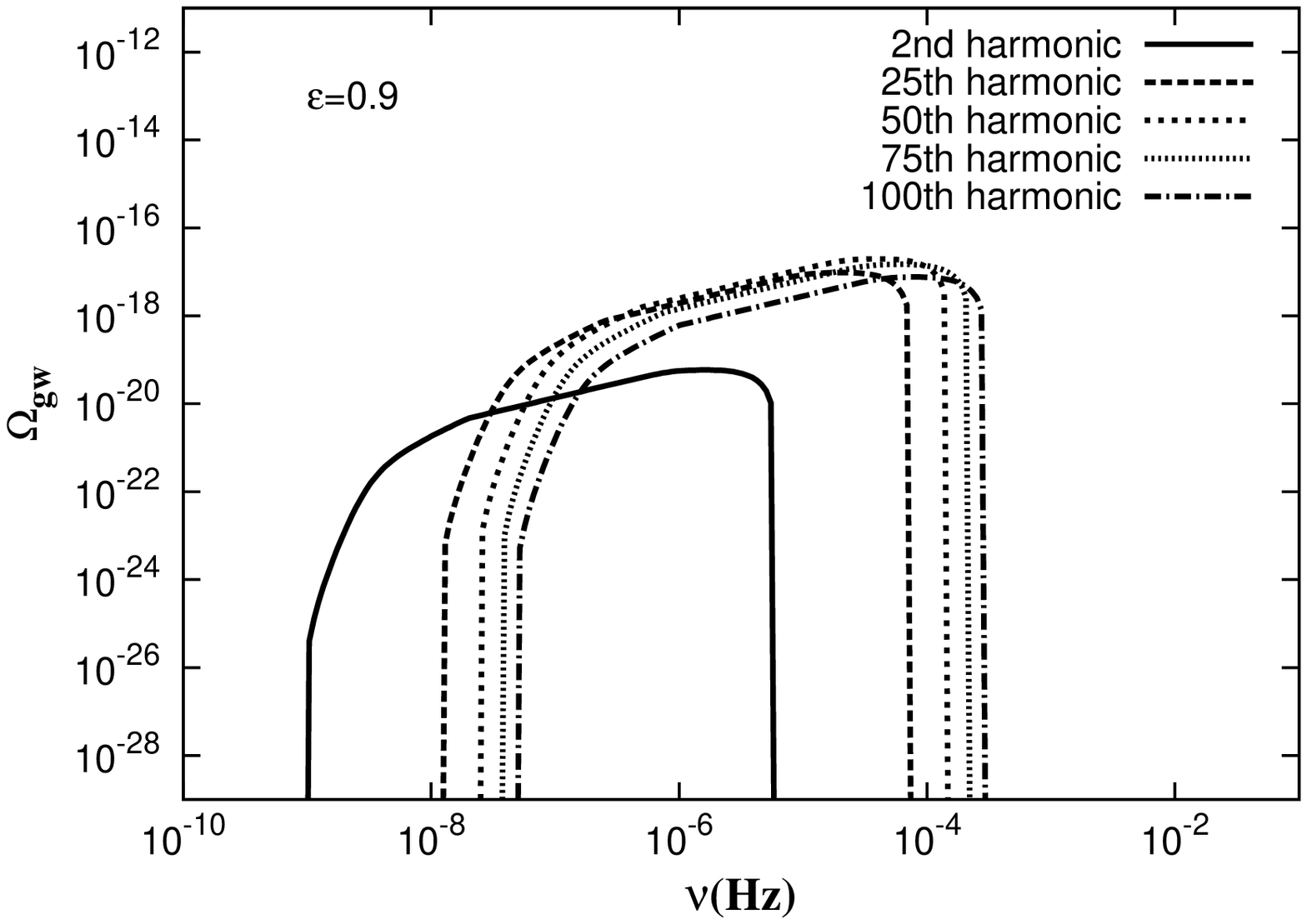}}
\caption{Same as in Fig. 1 for eccentricity $0.9$.}
\label{fig4}
\end{figure}

\begin{figure}
\centering
\rotatebox{-90}{\includegraphics[width=92mm]{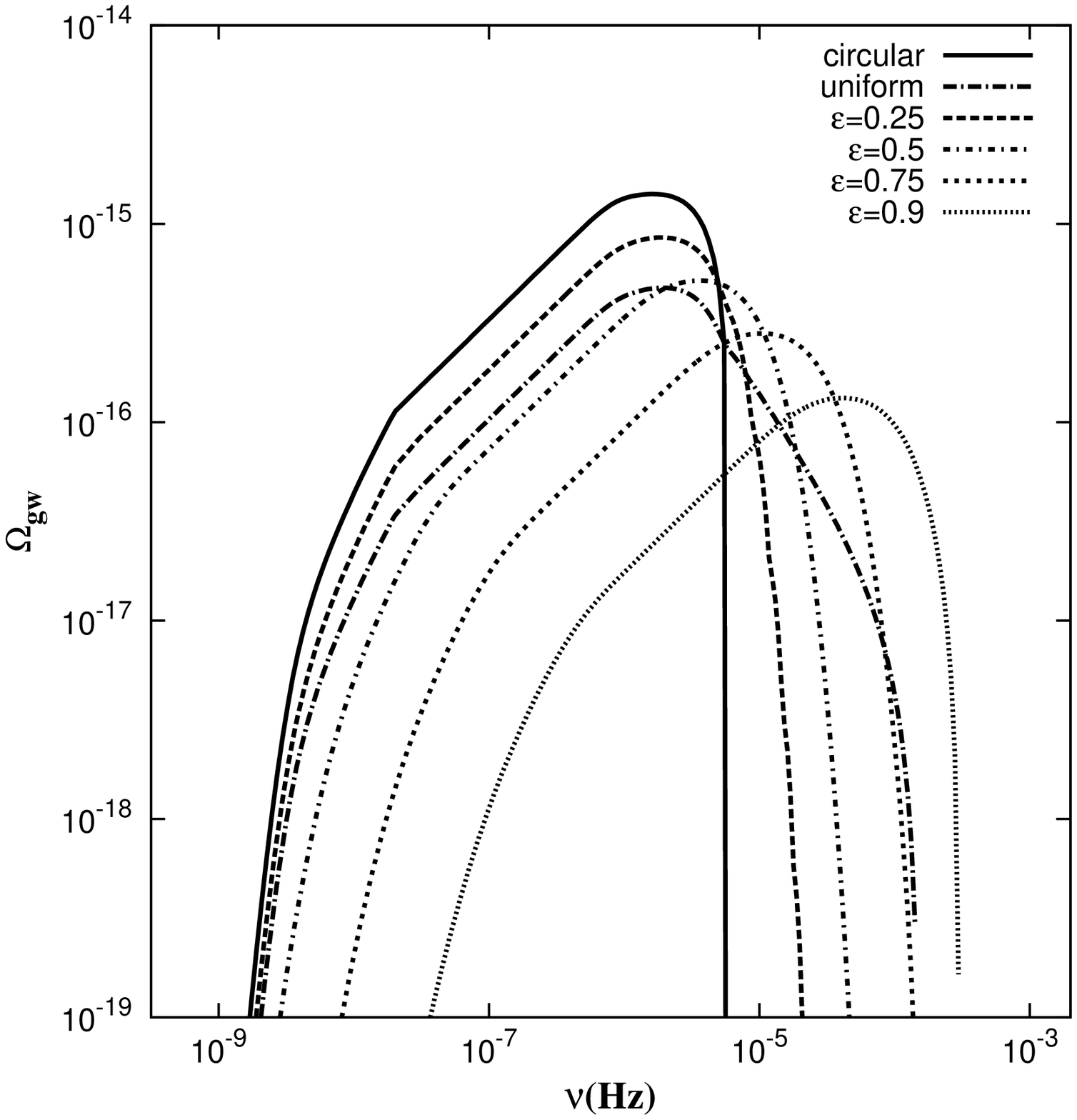}}
\caption{Total spectra (obtained by summing all relevant harmonics) for the cases corresponding to Figs~1-4. Also shown, the backgrounds generated by DNSs in circular orbits and by the uniform distribution of the eccentricities given by equation~(\ref{eq16}).}
\label{fig5}
\end{figure}

We show in Figs \ref{fig1}-\ref{fig4} the spectra generated by some harmonics of the emitted signals, considering the distribution given by equation~(\ref{deltaEccentr}). It is worth mentioning that, although we show in these figures only some harmonics, we include in the calculations of the total spectra (Fig.~\ref{fig5}) all relevant harmonics.

Note that for high eccentricities, the amplitudes corresponding to the higher harmonics become extremely relevant. Particularly, it is worth noticing the contrast between the cases corresponding to $\varepsilon=0.25$ and $\varepsilon=0.9$.

We show in Fig. ~\ref{fig5} the backgrounds generated when we sum the contributions of all relevant harmonics for the cases corresponding to Figs.~\ref{fig1}-\ref{fig4}. We also include in this very figure the spectra for: (a) the uniform distribution given by equation~(\ref{eq16}); (b)  DNS systems in circular orbits.

Comparing the \textit{uniform} curve in Fig. ~\ref{fig5} with, for example, that by a population of DNSs in circular orbits, we can observe the effects of the orbital eccentricities: the frequency band of the background is broadened and shifted towards higher frequencies.

The frequency band of the spectrum for circular systems ranges from $\sim 10^{-9}$ to $\sim 10^{-5}$Hz, whereas for the uniform distribution function the spectrum ranges from $\sim 10^{-9}$ to $\sim 10^{-4}$Hz.

Also, the amplitudes of the spectra are slightly reduced in the region between $\sim 10^{-9}$ and $\sim 10^{-5}$Hz whether the eccentricity of the population of DNSs is increased. This occurs due to the reduction of the amplitudes of the contributions corresponding to the lowest harmonics as compared to the highest ones.

This effect can be more clearly observed in the other spectra present in Fig.~\ref{fig5}, namely for $\varepsilon=0.25$, the maximum amplitude occurs at a frequency of $\sim 10^{-5}$Hz; whereas for $\varepsilon=0.9$ the maximum occurs at $\sim 10^{-4}$Hz. As expected, the higher the eccentricity is, the higher the frequency corresponding to the maximum amplitude of the spectrum. Moreover, the amplitudes have maximum values of $\sim 10^{-15}$ and $\sim 10^{-16}$ for $\varepsilon=0.25$ and $\varepsilon=0.9$, respectively.

A relevant issue is to investigate the detectability of the backgrounds studied here. Given the frequency bands involved, we consider the detectability by pulsar timing array (PTA) techniques and by the planned (space-based) antenna eLISA.

We refer the reader to, for example, \citet{moore}, where the sensitivity curves of several detectors, including the ones we mentioned here, are provided.

Concerning the PTA technique, the frequency band of operation ranges from $\sim 10^{-9}$ to $\sim 10^{-6}$Hz, with maximum sensitivity $\sim 10^{-9}$Hz. Particularly, in terms of the energy density parameter, EPTA (IPTA) has a maximum sensitivity of $\sim 10^{-10}$ ($\sim 10^{-12}$), whereas SKA is planned to have maximum sensitivity of $\sim 10^{-15}$.

The planned eLISA (and LISA, its precursor) has maximum sensitivity, in terms of the energy density parameter, of $\sim 10^{-12}$ ($\sim 10^{-10}$) around $\sim 10^{-3}$Hz. 

Given the sensitivity curves of PTA and eLISA (LISA), we conclude that the backgrounds shown in Fig.~\ref{fig5} would not be detected.

\section[]{Conclusions}
\label{sec7}

In this paper we show how to obtain the stochastic background of GWs generated by long-lived DNSs in eccentric orbits. We calculate the spectra by means of a different method: we determine the spectra that would be generated by each harmonic and then we sum these contributions in order to obtain the total backgrounds. In addition, we study the role of the eccentricity on the resulting spectra. This is achieved by using different distribution functions for the eccentricity. Particularly, we use two distribution functions: the Dirac delta function given by equation~(\ref{deltaEccentr}) and the uniform distribution given by equation~(\ref{eq16}).

From Fig.~\ref{fig5}, we note that the distribution of eccentricities affect the amplitudes and the frequency bands of the backgrounds. For example, we note that using the distribution given by equation~(\ref{eq16}), the amplitudes decrease in the frequency band ranging from  $\sim 10^{-9}$ to $\sim 10^{-5}$Hz, relatively to the case of a population of systems in circular orbits (see Fig.~\ref{fig5}). Concerning the frequency bands of the backgrounds, they are enlarged and shifted towards higher frequencies.

In general, we note that the higher the eccentricity of the population is the higher the frequency corresponding to the maximum amplitude of the spectrum. This behaviour is caused by the relative contribution given by the harmonics: for high eccentricities, higher harmonics affect more significantly the backgrounds.

Concerning the detectability of the spectra studied here, they would not be detected by the presently proposed space-based antennas, since the amplitudes involved are very small. Similar conclusion holds if we consider detections using the PTA technique.

In forthcoming papers we will address the study to systems formed by two black holes and those formed by a black hole and a neutron star. In addition, we will consider the high-frequency regimes, in which we cannot neglect the effects of the time evolution of the orbital parameters. Such cases are worth studying because, among other things, the spectra generated could form foregrounds for the planned space based antennas eLISA, DECIGO and BBO.

In addition, we expect that in the cases in which the time evolution of the parameters is considered, the form of the distribution $f(\varepsilon)$ will affect the results significantly, since in this case the frequency and the eccentricity will be related to each other.

\section*{Acknowledgements}

EFDE would like to thank the Brazilian agencies Capes, CNPq and FAPESP for financial support. JCNA would like to thank FAPESP and CNPq for the partial financial support. Last but not least, we thank the referee for the careful reading of the paper, the criticisms, and the very useful suggestions that greatly improved our paper.

\bsp

\label{lastpage}


\begin{thebibliography}{99}

\bibitem[\protect\citeauthoryear{Carroll \& Ostlie}{2007}]{carroll} Carroll B.W., Ostlie D.A., 2007, An Introduction to Modern Astrophysics, 2nd edn. Addison-Wesley, San Francisco, CA

\bibitem[\protect\citeauthoryear{Douglass \& Braginsky}{1979}]{thorne} Douglass D.H., Braginsky V.B., 1979, in Hawking S.W., Israel W., eds, General Relativity: An Einstein Centenary Survey. Cambridge Univ. Press, Cambridge, p. 99

\bibitem[\protect\citeauthoryear{Enoki \& Nagashima}{2007}]{enoki} Enoki M., Nagashima M., 2007, Prog. Theor. Phys., 117, 241

\bibitem[\protect\citeauthoryear{Ferrari et al.}{1999a}]{ferrari1} Ferrari V., Matarrese S. and Schneider R., 1999a, MNRAS, 303, 247

\bibitem[\protect\citeauthoryear{Ferrari et al.}{1999b}]{ferrari2} Ferrari V., Matarrese S. and Schneider R., 1999b, MNRAS, 303, 258

\bibitem[\protect\citeauthoryear{Ferrari et al.}{2001}]{ferrari3} Ferrari V., Matarrese S., Schneider R. and Portegies Zwart S.F., 2001, MNRAS, 324, 797

\bibitem[\protect\citeauthoryear{Heger et al.}{2003}]{heger} Heger A., Fryer C. L., Woosley S. E., Langer N. and Hartmann D. H., 2003, ApJ, 591, 288

\bibitem[\protect\citeauthoryear{Ignatiev et al.}{2001}]{ignatiev} Ignatiev V. B., Kuranov A. G., Postnov K. A., Prokhorov M. E., 2001, MNRAS, 327, 531

\bibitem[\protect\citeauthoryear{Kowaska et al.}{2012}]{kowalska2} Kowalska I., Bulik T., Belczynski K., 2012, A\&A, 541, A120

\bibitem[\protect\citeauthoryear{Moore et al.}{2015}]{moore} Moore C. J., Cole R. H., Berry C. P. L., 2015, Class. Quantum Gravity, 32, 015014

\bibitem[\protect\citeauthoryear{Peters}{1964}]{peters2} Peters P.C., 1964, Phys. Rev., 136, B1224

\bibitem[\protect\citeauthoryear{Peters \& Mathews}{1963}]{peters} Peters P.C., Mathews J., 1963, Phys. Rev., 131, 435

\bibitem[\protect\citeauthoryear{Regimbau}{2011}]{regimbau2} Regimbau T., 2011, Res. Astron. Astrophys., 11, 369

\bibitem[\protect\citeauthoryear{Regimbau \& Chauvineau}{2007}]{regimbau1} Regimbau T. and Chauvineau B., 2007, Class. Quantum Gravity, 24, S627

\bibitem[\protect\citeauthoryear{Regimbau \& Pacheco}{2006}]{regimbau3} Regimbau T., de Freitas Pacheco J.A., 2006, ApJ, 642, 455

\bibitem[\protect\citeauthoryear{Rosado}{2011}]{rosado} Rosado P.A., 2011, Phys. Rev. D, 84, 084004

\bibitem[\protect\citeauthoryear{Salpeter}{1955}]{salpeter} Salpeter E.E., 1955, ApJ, 121, 161

\bibitem[\protect\citeauthoryear{Smartt}{2009}]{smartt} Smartt S.J., 2009, ARA\&A, 47, 63

\bibitem[\protect\citeauthoryear{Springel \& Hernquist}{2003}]{sfrd} Springel V. and Hernquist L., 2003, MNRAS, 339, 312

\end{thebibliography}
\end{document}